\documentclass[10pt,twocolumn,letterpaper]{article}

\usepackage[pagenumbers]{cvpr} 

\usepackage[dvipsnames]{xcolor}

\definecolor{cvprblue}{rgb}{0.21,0.49,0.74}
\usepackage{graphicx}
\usepackage{nicematrix,enumitem}
\usepackage{booktabs}
\usepackage{multirow}
\usepackage[tableposition=top]{caption}
\usepackage{subcaption}
\usepackage[pagebackref,breaklinks,colorlinks,citecolor=cvprblue]{hyperref}

\def\paperID{39} 
\def\confName{CVPR}
\def\confYear{2024}

\title{Using Multiparametric MRI with Optimized Synthetic Correlated Diffusion Imaging to Enhance Breast Cancer Pathologic Complete Response Prediction}

\author{Chi-en Amy Tai \\
University of Waterloo\\
Waterloo, ON\\
{\tt\small amy.tai@uwaterloo.ca}
\and Alexander Wong \\
University of Waterloo\\
Waterloo, ON\\
{\tt\small alexander.wong@uwaterloo.ca}
}

\begin{document}
\maketitle
\begin{abstract}
In 2020, 685,000 deaths across the world were attributed to breast cancer, underscoring the critical need for innovative and effective breast cancer treatment. Neoadjuvant chemotherapy has recently gained popularity as a promising treatment strategy for breast cancer, attributed to its efficacy in shrinking large tumors and leading to pathologic complete response. However, the current process to recommend neoadjuvant chemotherapy relies on the subjective evaluation of medical experts which contain inherent biases and significant uncertainty. A recent study, utilizing volumetric deep radiomic features extracted from synthetic correlated diffusion imaging (CDI\textsuperscript{s}), demonstrated significant potential in noninvasive breast cancer pathologic complete response prediction. Inspired by the positive outcomes of optimizing CDI\textsuperscript{s} for prostate cancer delineation, this research investigates the application of optimized CDI\textsuperscript{s} to enhance breast cancer pathologic complete response prediction. Using multiparametric MRI that fuses optimized CDI\textsuperscript{s} with diffusion-weighted imaging (DWI), we obtain a leave-one-out cross-validation accuracy of 93.28\%, over 5.5\% higher than that previously reported.  
\end{abstract}    
\section{Introduction}
\label{sec:intro}
In 2020, 685,000 deaths across the world were attributed to breast cancer, underscoring the critical need for innovative and effective breast cancer treatment~\cite{bca-who-stats}. Recently, neoadjuvant chemotherapy emerged as a promising treatment strategy, gaining popularity due to its ability to shrink large tumours~\cite{survival-bc-patients} and elicit a pathologic complete response (pCR) for breast cancer patients~\cite{pcr-bc}. Unfortunately, neoadjuvant chemotherapy presents challenges in terms of cost, time commitment, and potential risks for patients, including exposure to radiation and the possibility of encountering significant side effects such as decreased fertility~\cite{cost-effectiveness-nac}. 

The current methodology for recommending neoadjuvant chemotherapy hinges on the subjective evaluation by medical experts, assessing the patient's prognosis and potential treatment benefits~\cite{nac-human-recommendation}. The inherent biases and significant uncertainty in human judgment introduce the possibility of misguided recommendations~\cite{human-judgement-problem}, putting some patients at risk of preventable progression to advanced cancer or unnecessary exposure to radiation. Advances in computer vision and medical imaging have demonstrated the potential for precise and noninvasive techniques in predicting pathologic complete response prediction~\cite{tai2023enhancing}.

Most promising, using a pretrained 34-layer volumetric residual convolutional neural network architecture with initial weights from MONAI~\cite{monai} with synthetic correlated diffusion MRI images achieved a leave-one-out cross-validation (LOOCV) accuracy of 87.75\%. Initially introduced in~\cite{wong2022synthetic}, CDI\textsuperscript{s} is a derived MRI modality that combines native and synthetic diffusion signal acquisitions. This combination, along with signal calibration, aims to enhance consistency for the dynamic range across different machines and protocols. The study in~\cite{wong2022synthetic} illustrates the effects of fine-tuning CDI\textsuperscript{s} for prostate cancer. Conversely, in~\cite{tai2023enhancing}, modifications to CDI\textsuperscript{s} for breast cancer were minimal, lacking proper tuning of the coefficients. Inspired by the positive outcomes of optimizing CDI\textsuperscript{s} for prostate cancer delineation~\cite{wong2022synthetic}, this paper investigates the application of optimized CDI\textsuperscript{s} to enhance the prediction of breast cancer pathologic complete response. 

\section{Methodology}
\label{sec:methodology}
The data is obtained from the pre-treatment (T0) cohort in the American College of Radiology Imaging Network (ACRIN) 6698/I-SPY2 study~\cite{acrin6698-data-1, acrin6698-data-2, acrin6698-data-3, acrin6698-data-4} filtered for non-null pCR values, leading to a total of 253 remaining patients. As the classes were imbalanced (class 0:1 is 67.6\%:32.4\%), a weighted random sampler was added to the training sampler along with an AdamW optimizer. In addition, a cosine annealing learning rate scheduler was also implemented during training.

\begin{figure*}
    \centering
    \includegraphics[width=\linewidth]{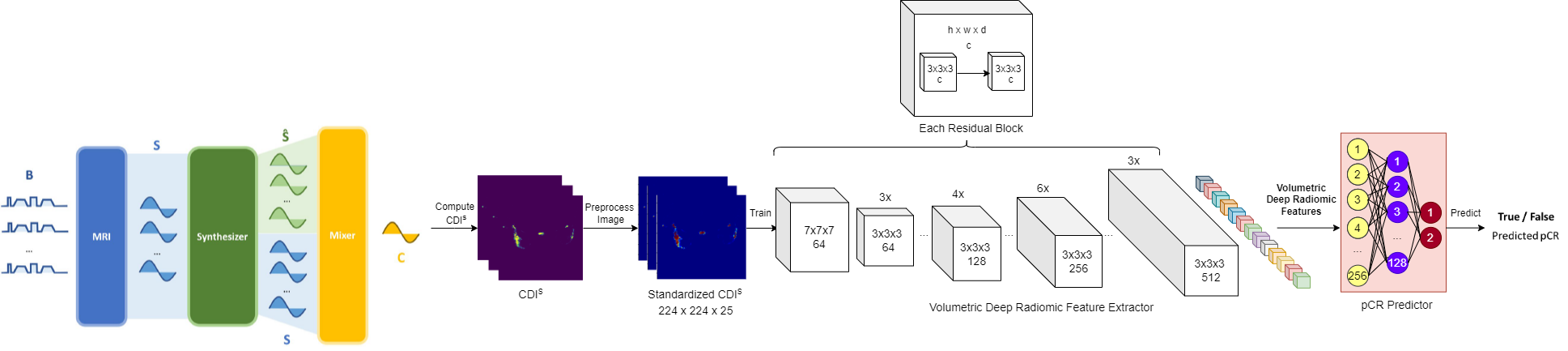}
    \caption{Clinical support workflow copied from~\cite{tai2023enhancing} that is used in this study.}
    \label{fig:support-workflow}
\end{figure*}

To optimize CDI\textsuperscript{s}, the Nelder-Mead simplex optimization strategy was used to maximize the area under the receiving operating characteristic curve (AUC) for breast cancer tumour delineation. Then, the CDI\textsuperscript{s} signals were fused with diffusion-weighted imaging (DWI) to create a multiparametric MRI. Finally, all patient volumes were then standardized to 224x224x25 volumetric data cubes for dimensional consistency.

Building on the foundation of a previously developed deep radiomic clinical support system outlined in~\cite{tai2023enhancing} (see Figure~\ref{fig:support-workflow}), we adopt a 34-layer pre-trained volumetric residual convolutional neural network framework from MONAI to obtain deep radiomic features. The weights from MONAI were obtained from training on eight different 3D MRI and CT segmentation datasets~\cite{chen2019med3d}. The radiomic features were subsequently inputted into a fully-connected neural network predictor, designed to classify breast cancer into no pCR or pCR. 

For training, a learning rate of 0.001 was also used. All the model layers were trained with no freezing and leave-one-out cross-validation was used to compare the performance. Though the previous paper only reported the average accuracy, we also provide the average sensitivity, specificity, and F1 score. 

\section{Results}
\label{sec:results}
\begin{figure}
    \centering
    \subfloat[]{\includegraphics[width=0.48\linewidth]{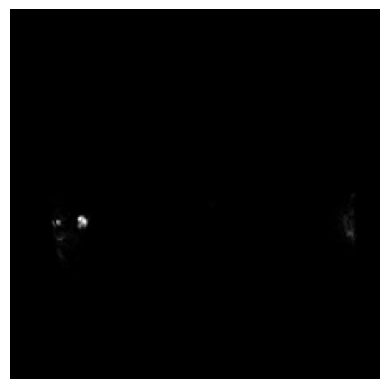}} 
    \subfloat[]{\includegraphics[width=0.48\linewidth]{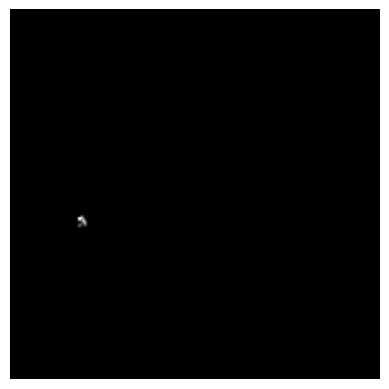}} \\
    \subfloat[]{\includegraphics[width=0.48\linewidth]{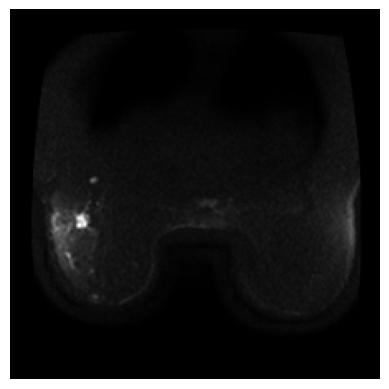}} 
    \subfloat[]{\includegraphics[width=0.48\linewidth]{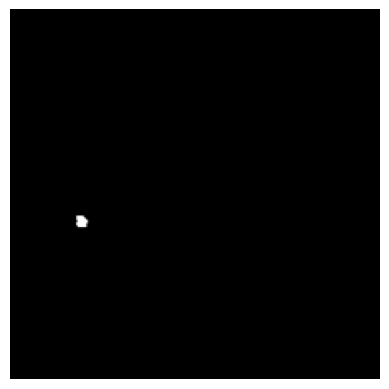}} 
    \caption{An example slice illustrating visual differences between (a) Unoptimized CDI\textsuperscript{s}, (b) Optimized CDI\textsuperscript{s}, (c) the associated DWI image, and (d) the associated tumour mask at pre-treatment for a patient who obtained pCR. For this patient, the pCR prediction was correct using the Optimized CDI\textsuperscript{s} signal fused with DWI.}
    \label{fig:v2-grade-comparison-results}
\end{figure}

Table~\ref{tab:pcr-modalities-monai-resnet-v2} shows the results from using the unoptimized and optimized CDI\textsuperscript{s} with DWI to create a multiparametric MRI. As shown, the accuracy using the optimized CDI\textsuperscript{s} was 93.28\%, over 5.5\% higher than that previously reported (87.75\%). Though the sensitivity and specificity metrics were not previously reported, using optimized CDI\textsuperscript{s} achieved over 90\% for both sensitivity and specificity. Notably the F1 score using optimized CDI\textsuperscript{s} was also high at 90.17\%. An example highlighting the visual differences between the imaging modalities of the unoptimized CDI\textsuperscript{s} and optimized CDI\textsuperscript{s} are shown in Figure~\ref{fig:v2-grade-comparison-results}. Also provided in Figure~\ref{fig:v2-grade-comparison-results} are the associated tumour mask and DWI for the specific patient slice.

\begin{table}[ht]
    \caption{Results using unoptimized and optimized CDI\textsuperscript{s} with the best result bolded.}
    \centering
    \begin{NiceTabular}{l c c c}
        \toprule
        \RowStyle{\bfseries}
        CDI\textsuperscript{s} Version & Accuracy & Sensitivity & Specificity \\ \midrule
        Unoptimized & 87.75\% & N/A & N/A \\
        Optimized & \textbf{93.28\%} & \textbf{95.12\%} & \textbf{92.40\%} \\
    \bottomrule
    \end{NiceTabular}
    \label{tab:pcr-modalities-monai-resnet-v2}
\end{table} 

Given the promising results, this proposed noninvasive method that predicts pathologic complete response would enhance patient treatment with minimal side effects (as it uses MRI modalities that are normally obtained in the course of the diagnosis). In addition, the results highlight the importance of tuning CDI\textsuperscript{s} for the specific cancer domain as the optimized CDI\textsuperscript{s} modality obtains superior performance compared to unoptimized CDI\textsuperscript{s}.

{
    \small
    \bibliographystyle{ieeenat_fullname}
    \bibliography{main}
}

\end{document}